# A DEVELOPMENT OF HYBRID FRAMEWORK FOR E-GOVERNMENT

**Ahmed Mateen**[*]

**Sana Sabir***

**Kareem Ullah***

**Abstract**

Governments all around the world are widely investing on the implementation of e government to advance services to citizens and minimize costs. Governments can progress effectiveness of their operations and can carry their administrative operations efficiently with the help of ICT. Electronic government perceived to provide a way for governments to renovate their operational activities to serve their clients more competently. With improvement in Information and Communication Technology (ICT), it is now time to device electronic access to government facilities to the variously located citizens. E governments all around the world have different objectives and follow different models for e government development. Present models examined and found less than satisfactory to guide e-government implementation. This research proposed a hybrid model from "Citizen comprehensive vision acknowledged the civic idea" and "The strategic framework of e-government" models. The procedure of merging different computational knowledge systems to assemble a solitary crossover show has turned out to be progressively prominent. The execution files of these mixture models have ended up being superior to the individual segments when utilized alone. To ensure that the proposed model is more efficient and beneficial a survey study conducted. Survey based on questionnaires and results calculated by applying statistics on data gathered from survey.

[*] **Lecturer Computer Science, University of Agriculture Faisalabad**

# 1. Introduction

E-government implies the expending of Information Communication Technologies (ICT) uniquely the web to bring the data and administrations of subjects various government advancement models of e-government seen at back-to-back time by worldwide associations and logical foundations and a few analysts. These past models incorporate diverse successive procedural strategies that characterize the procedure of selection and advancement of the e-government from various perspectives as indicated by rising prerequisites [1]. E-Government in its least difficult structure can mean utilizing data and correspondence innovation apparatuses to give administrations to nationals [2].

E governments all around the world have different objectives and follow different models for e-government development that included new ideas, for example, straightforwardness, responsibility, native interest in the assessment of government execution [3]. It exhibited capacities and consequently distinguishes obstructions inside e-Government usage in Pakistan [4].

Generally, there are four stages of e government as shown in Figure 1.
E-government offers administrations to those inside of its power to execute electronically with the legislature. These administrations contrast as per clients' needs, and this differences have offered ascend to the advancement of various kind of e-government. E-government capacities ordered into four principle classifications. In order to propose a hybrid model, two models selected the first model studied was "citizen comprehensive vision acknowledged through ID card Integrated Delivery of E-government Applications" [5]. It concentrated creating basic and successful devices and methodologies for overseeing, understanding and executing e government activities. Second model was "The strategic framework of e-government" [6].

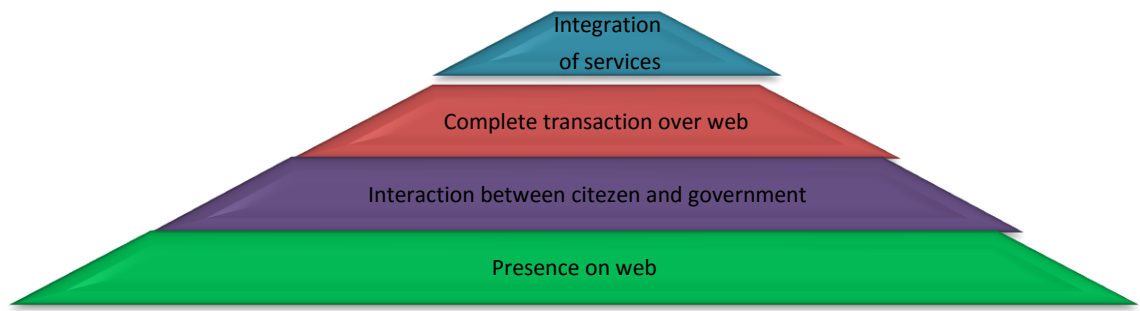

Figure 1. Stages of e-Governance

This model is the consequence of survey of e government techniques of 20 nations notwithstanding European Union. The usage of the model additionally gave a few suggestions to policymakers at the national and organization levels [7].

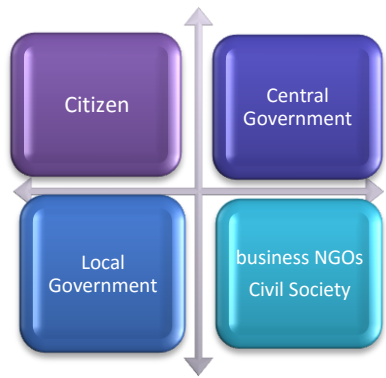

Figure 2. Types of e-Government Application

In applying the considered G2C, customers have minute and supportive access to government information and organizations from everywhere at whatever time, by method for the use of different channels. Despite making certain trades, for instance, certifications, paying authoritative costs, and applying for focal point.Government to business, or G2B, is the second genuine kind of e-government class. G2B can pass on essential efficiencies to both governments and associations. G2B fuse diverse organizations exchanged the center of government and the business parts, including scattering of methodologies, redesigns, standards and directions. This system focal points government from business online experiences in extents, for instance, e-showcasing methods. The organization to-business G2B is as useful as the G2C system, redesigning the capability and nature of correspondence and trades with business in like manner, it constructs the value and straightforwardness of government contracting and augments. The key purpose of G2G headway is to redesign and upgrade between government definitive techniques by streamlining investment and coordination. To build up a model for executing neighborhood e-government is a case of a creating nation [8].

## 2. Previous Work

Eleven e-government development models examined and dissect (eGMMs) for security administrations. Facilitate, it endeavors to build up a typical casing of reference for eGMM basic stages. The study used the Soft Systems Methodology (SSM) of logical request/taking in cycle received from Check and Scholes [9]. E-government ventures confront part of difficulties that may prompt to disappointment of e-government. Most vital of these difficulties incorporate the HR are not qualified and the high cost of framework and the conventional foundation can't scale, versatility requests change after some time. The arrangement of these difficulties distributed, computing to cost sparing, proficient administration of assets and applications and augmentation in the capacity of adaptability [10].

In view of the writing, it guaranteed that the accessibility of a successful e-Government evaluation structure is a vital condition for propelling e-Government appropriate usage [11]. The Effective e-government is turning into an essential go for some legislatures around the globe [12]. E-Government security viewed as one of the vital elements for accomplishing a propelled phase of e-government [13]. A comprehension of the data security innovation and the requirement for its usage is more secure and smooth working of e-administration undertaking [14].E-government idea explores CSFs for e-government by surveying the writing, and recognizes far reaching set of CSFs by activity research, conceptualizing and the Delhi ponder. Finally, it proposes the model of CSFs [15].

The rise of Electronic Government as a device of practice in the re-innovation of Governance presents e-Government as far reaching of Electronic Democracy, and Electronic Business, inspects the nature and extent of improvements in this rising field [16]. E-government security viewed as one of the urgent elements for accomplishing a propelled phase of e-government [17].A half-and-half model proposed from different creators who have attempted to demonstrate the execution of e-government. In this short paper the creators exhibit a brief contextual analysis of one example of open information discharge, concentrating on a dataset identified with the 'Advanced Landscape Research' distributed in 2012 nearby another Government Digital Strategy [18]. Web data comfort and online value-based offices amplify clearness, openness of bureaucratic foundations and thusly obligation and reduction expense of exchanges [19].

*A)   Problem Statement*

"The CIVIC" and "The strategic framework of e-government" models where seen best practice models in overall e-government models. Despite of their good characteristics these models still unable to fulfill overall requirements because they used alone. To enhance their abilities this research is used to conjunct their good characteristics to build up a new Hybrid model. The proposed model is more efficient and beneficial.

*B)   Objective*

The objective of this research is to develop a hybrid model from the "Citizen comprehensive vision acknowledge" and "The strategic framework of e-government" which is more efficient, easy to implement and beneficial for the smooth running of e government system.

**3. ResearchMethedology**

To propose best practice model, this research taking the following stages firstly, a review of literature conducted for both selected models to identify development of best practice in e-government. Review includes three phases that are Plan, Conduct, and Document as shown in Fig 3.

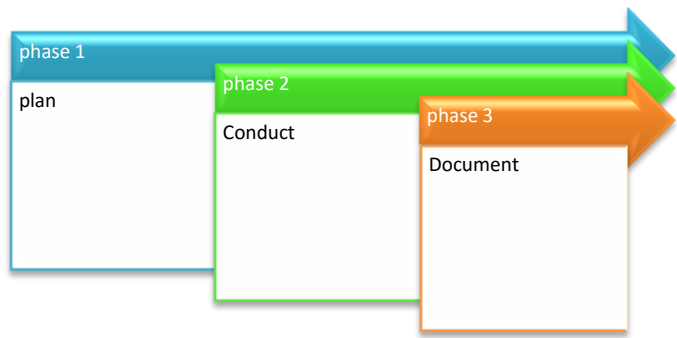

Figure 3. Conducting Existing Work

A comprehensive set of search terms in order to cover broad topics addressed by researches. Thus, the following sets of applied keywords, derived from preliminary searches and enriched by keywords found in literature, as the basis for our searches (e-government, hybrid model, best practice models in governance strategic frame for e governance). As far as points, high significance attached to responsibility/administration, external reporting, performance measurement, and planning in which the reception of accounting frameworks and their value is regularly discussed [20].The availability of a productive e-Government assessment structure is a key condition for impelling e-Government suitable use [21]. A substantial number of models and systems proposed for the compelling advancement and advancement of e-administration [22].

Secondly, based on literature finding, empirical observations and concerning e-government best practice model with their advantages and disadvantages of each phase of the models as shown in fig 4. Specifically, researchers performing systematic review must make effort to distinguish and report research that does not support their favored research hypothesis and recognizing and reporting research that backings it [23].

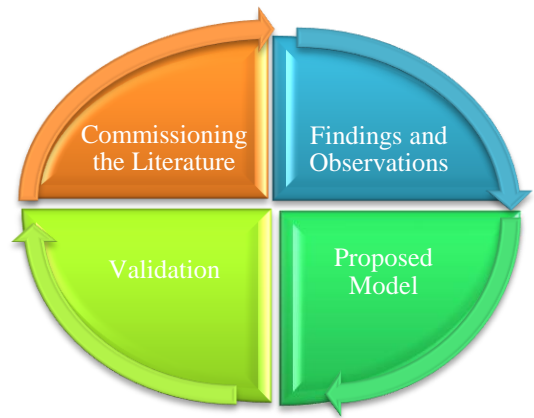

Figure 4. Methodology Overview

Thirdly, to prove that the proposed model is more efficient, cost effective, reliable and beneficial a survey conducted using questionnaires. For survey, a few inquiries adjusted and others added to reflect all the measuring builds up that exist in the proposed framework. The researches variables measured in a 7-point Likert's scale, with '1' as strongly disagree to the '7' as strongly agree. The number of people in the study was the employees in different departments of government sector where e-government is applied. There are averages 57 to 60 employees are involve e-government activities in different government department at district level. For the calculation of sample size of this study, Taro Yamane's formula been applied. The particulars are as under: $n=N/1+N(E)^2$ which is $57/1+57(0.05)^2=50$ , Where, n sample size, N total population

'e' margin of error (0.05) .The estimated sample size for this study was 50.By applying systematic sampling technique, the interval had been taken. *Interval (I)* = N/n =57/50 First respondent selected by applying simple random technique among first respondents. Remaining respondents selected by adding sampling interval of one until the completion of the sample size. The most vital specialty of statistical work is the data accumulation. The information gathered by the researcher himself in a face-to-face circumstance. Every one of the respondents interviewed personally.

*A) Statistical Techniques*

i) Descriptive statistics, counting frequencies and percentages used to compress diverse factors. To describe the variables, Simple frequency tables constructed out of data. To find out the frequency distribution of the variables, simple percentage was calculated. The percentages calculated by following formula: p=F/N *100 where P is Percentage, F is Frequency and N is Total Number of frequencies ii). Inferential statistics and Pearson correlation used to evaluate the relationship among dependent and independent variables. Bivariate correlation test applied to test the relationship among dependent and independent variables. It shows how much strongly or weakly independent and dependent variables are associated. Formula for calculation of correlation is as follows:

$$r = \frac{n(\Sigma xy) - (\Sigma x)(\Sigma y)}{\sqrt{[n\Sigma x^2 - (\Sigma x)^2][n\Sigma y^2 - (\Sigma y)^2]}}$$

In light of the outright significance of the key targets behind e-government usage, they highlighted in the national e-government technique. They alluded to as vital here in light of the fact that they come from the e-government methodology. It is straightforwardness basic for these targets dependably are an intelligible part of any e-government activity. It is critical to have an applied establishment that passes on the different segments of the arrangement. In doing visualize e-government through a water stream display. In such a model, the overall elected e-government methodology needs to fathom by elected and nearby offices thusly make an interpretation of these systems into e-administrations for the nationals.

The stage includes following entities:

- Vision
- objective
- Guiding principles
- Strategy (transform strategy , convey details)
- Support services
- Share services
- Enterprise services
- Channels
- Delivery channels (G2C,G2B)
- Integration channels (portals)

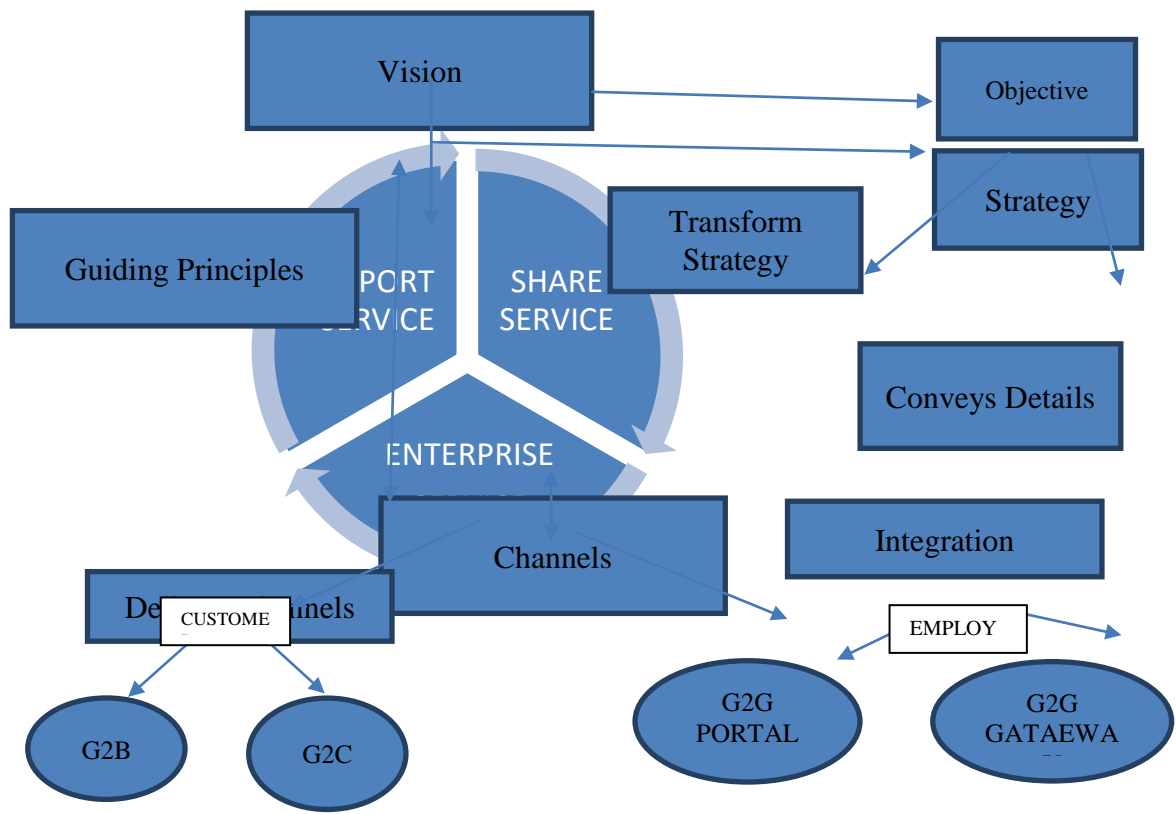

Figure 5. Proposed Model

- 

## B) Why This Method is Used

This methodology used to recognize comparative work done inside the territory and distinguishing information gaps that request further examination and to look at past discoveries.

Rapidly uncover which researchers have composed the most on a specific subject and are, hence, presumably the specialists on the theme. As a rule, an analyst may find new edges that need encourage investigation by inspecting what as of now been composed on a subject. It is regularly valuable to audit the sorts of studies that past scientists have propelled as a method for figuring out what methodologies may be of most advantage in further building up a subject Surveys comes about give a preview of the states of mind and practices including considerations, feelings, and remarks about target overview populace.

This profitable criticism is your standard to scale and build up a benchmark from which to look at results after some time.The structured interviews used since it produces quantitative information, taking after the inspecting system and the methodologies utilized as a part of standard reviews. To determine the effectiveness of interview schedule, it is necessary to pre-test it before actually using it. Pretesting can help to determine the strengths and weaknesses of survey concerning question format, wording and order. Interview schedule pre-tested on 10 respondents before starting the actual research.

## 4. Results

A few inquiries changed and others added to mirror all the measuring develops that exist in the recommended system. Table 1 shows the questions in our study the research variables are measured with a seven-point scale, where 1 represents strongly disagree, and 7 as strongly agree. Result calculated over responses for each question as shown in table 1.

Increases or decreases in one variable do significantly relate to increases or decreases in second variable. Results depicts that our model has provides better services for E-governance in different departments of e-government. Figure 6 shows level of satisfaction and adoption by the respondents.

The research variables are measured in a scale, which indicates model adoption was more than 10% strongly agreed, and the average of agree responses 60% also almost 28% responses was over model adaption. According to measured variables and responses over model satisfaction, the percentages of weakly agree, agree and neutral are 22, 70 and 8 respectively. Results point to the better services of e government in different departments.

Table 1.E-government Adoptation summary of the proposed Model

| Constructs | Dimension | Response % | | | | | | |
|---|---|---|---|---|---|---|---|---|
| | | **Strongly Disagree** | **Disagree** | **Weakly Disagree** | **Neutral (N)** | **Weakly Agree** | **Agree** | **Strongly Agree** |
| **Strategy:** This section identifies the main drivers for the implementing e-government. | Strategy Definition and Internet Activities | | | | 26% 20% | 52% 60% | 22% 20% | |
| | Responsibilities and Roll of Stakeholder | | | | 40% 39% | 48% 49% | 12% 14% | |
| | Model Collaboration with Agencies | | | | 22% | 58% | 20% | |
| **Processes:** This section measures the perceptions toward the processes of model. | Business Change and Defined Process | | | | 28% 28% | 42% 54% | 30% 16% | 2% |
| | Business Documenting and Integration | | | | 32% 30% | 40% 40% | 28% 30% | |
| | Monitoring Evaluation and Impact assessment | | | | 40% 24% | 44% 54% | 16% 22% | |
| **Technology:** It compares IS structures, covers information, system and web presence quality also data protection and integration. | Data Sharing in Organization And Data Sharing with other Agencies | | | | 24% 24% | 38% 48% | 38% 28% | |
| | Internet Applications and Integration of Applications | | | | 28% 32% | 52% 40% | 20% 28% | |
| | Legacy Systems and | | | | 32% 24% | 40% 54% | 28% 22% | |
| | Two Way Interaction | | | | | | | |
| **People:** Several constructs provided in this dimension such as user and employee satisfaction, impact adaptation, and also overall feedback of the respondent. | Personnel Satisfaction and Performance Impact | | | | 08% 08% | 22% 22% | 70% 54% | 16% |
| | Simplicity in Adoption and Communication | | 28% | 14% | | 60% 72% | 12% 14% | |
| | Services providing and Personnel likeness | | | | 20% 32% | 48% 56% | 32% 08% | 04% |

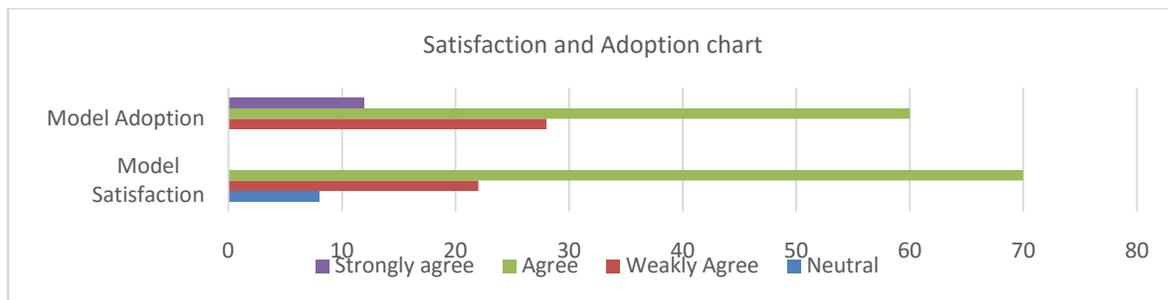

Figure 6. Model satisfaction and adoption

## 5. Conclusion

This research proposed a hybrid model from "Citizen comprehensive vision acknowledged" and "The strategic framework of e-government" models who have attempted to demonstrate the improvement of best practice in any e government activity. Survey writing was behavior to grow new model. Keeping in mind the end goal to guarantee the proposed model proficiency and advantages, a study directed. The recommended model surveyed e-government in an open association covering every single inner variable. It ordered these elements into four primary measurements: strategy, processes, technology, and people. Various estimation develops proposed under every measurement. The primary phase of testing the structure performed through directing review on government associations in Pakistan by getting its representatives' observation diverse perspectives in their association. The study analyzed the heaviness of each of the four measurements in e-government and the connections between them. Further studies on extra cases are occurring. The study discoveries affirmed the exploration theories demonstrating that every one of the four measurements influence e-government model execution however with various weights. Comes about likewise demonstrated that e-Government information sharing does not influence the procedure of e-Government forms with the association's business forms. As a conclusion, the examination prescribes that in considering unmistakable e-Government endeavors and activities, one ought to consider all inside e-Government building pieces: system, methods, advancement, and people.

## 6. Future Work

To make e government more productive and gainful, two technologies can shelter to it. One of them is open source software (OSS), second is cloud computing. The merge of OSS with e government is most alluring in developing countries as they provide various cost benefits, expanded adaptability to address localization issues and extensibility, increased ownership and local autonomy. Cloud computing is no doubt one of the most life changing technique in IT range, this could be used in e government to boost up the outcomes of e government.

## References

[1]     Alhomod, S. M. and M. M. Shafi, 2012. Best Practices in E government: A review of Some Innovative Models Proposed in Different Countries. *International Journal of Electrical and Computer Sciences IJECS-IJENS*, *12*.

[2]     Al-shboul, M., Rababah, O., Al-shboul, M., and R. Ghnemat, 2014. Challenges and Factors Affecting the Implementation of E-Government in Jordan, (December), 1111–1127.

[3]     Mohammad, H., T. Almarabeh and A. A. Ali, 2009. E-government in Jordan. *European Journal of Scientific Research,* 35(2): 188-197.

[4]     Kayani, M. B., M. E. Haq, M. R. Perwez and H. Humayun, 2011. Analyzing Barriers in e-Government Implementation in Pakistan. *International Journal*, *4*(3): 494–500.

[5]     Al Khouri, A. M. (2011). "An Innovative Approach for E-Government Transformation". *International Journal of Managing Value and Supply Chains*, *2*(1): 22–43.

[6] Rabaiah, A. and E. Vandijck, 2009. "A Strategic Framework of e-Government". *Generic and Best Practice,7*(3): 241–258.

[7] Valdés, G., M. Solar, H. Astudillo, M. Iribarren, G. Concha and M. Visconti, 2011. Conception, Development and Implementation of an E-Government Maturity Model in Public Agencies. *Government Information Quarterly*, 28(2): 176-187.

[8] Nabafu, R. and G. Maiga, 2011. Towards A Model for Implementing Local E-Government in Uganda *International Journal of Reviews in Computing*, 286–306.

[9] Karokola, G., L. Yngström and S. Kowalski, 2012. "Secure E-Government Services: A Comparative Analysis of E-Government Maturity Models for the Developing Regions–the Need for Security Services". *International Journal of Electronic Government Research (IJEGR),* 8(1): 1-25.

[10] Ali *et al*., 2012. "Proposed Development Model of E-Government to Appropriate Cloud Computing". *International Journal of Reviews in Computing*, 9(4): 47-53.

[11] Azab, N., S. Kamel and G. Dafoulas, 2009. A suggested framework for assessing electronic government readiness in Egypt. *Electronic Journal of E-Government*, *7*(1): 11–28.

[12] Alshehri, M. and S. Drew, 2010. E-Government Fundamentals. *Proceedings of the IADIS International Conference on ICT, Society and Human Beings*, 35–42.

[13] Omer, N., F. Elssied, O. Ibrahim and A. Yousif, 2011. Review Paper: Security in E-government Using Fuzzy Methods. *International Journal of Advanced Science and Technology,37*: 99–112.

[14] Singh, S. and D. Karaulia, 2011. E-Governance. *Information Security Issues on Computer Science and Information,* 1(9): 4-11.

[15] Case, T. H. E. and O. F. Poland, 2013." A Model of Success Factors for E-Government Adoption". *Issues in Information Systems*, 14(2): 87–100.

[16] Ali, M. and N. Mujahid, 2015. Electronic Government Re-Inventing Governance: A Case Study of Pakistan. *Public Policy and Administration Research*, 5(2): 1–9.

[17] Upadhyaya, P., S. Shakya and M. Pokhare, 2012. Information Security Framework for E-Government Implementation in Nepal. *Journal of Emerging Trends in Computing and Information Sciences*, *3*(7): 1074–1078.

[18] Daniel, M. 2015. "Electronic government: A Development Model for Papua New Guinea". *Divine Word University Research Journal*, 22(5):17–32.

[19] Kachwamba, M. and A. Hussein, 2009. Determinants of e-Government Maturity: Do Organizational Specific Factors Matter". *Journal of US-China Public Administration*, *6*(7): 1-8.

[20] Schmidt, U. and T. Günther, 2016. Public sector accounting research in the higher education sector: a systematic literature review. *Management Review Quarterly*, 3(1): 1-31.

[21] Azab, N., S. Kamel and G. Dafoulas, 2009. A suggested framework for assessing electronic government readiness in Egypt. *Electronic Journal of E-Government*, *7*(1): 11–28.

[22] Goel, I. 2013. A Characteristics Study of Existing Models for the Effective Development and Promotion of E-Governance. *International Journal of Advanced Research in Computer Science and Software Engineering,3*(1): 5–13.

[23] Morris, L., S. Wilson and W. Kelly, 2016. Methods of conducting effective outreach to private well owners–a literature review and model approach. *Journal of water and health*, *14*(2): 167-182.